\documentstyle[preprint,aps]{revtex}
\begin{document}

\title{Electronic-structure-induced deformations of liquid metal clusters}

\author{H. H\"akkinen and M. Manninen}

\address{Department of Physics, University of Jyv\"askyl\"a,
P.O.Box 35, FIN-40351 Jyv\"askyl\"a, Finland}

\date{\today}
\maketitle

\begin{abstract}
Ab initio molecular dynamics is used to study deformations of
sodium clusters at temperatures $500\cdots 1100$ K. Open-shell Na$_{14}$
cluster has two shape isomers, prolate and oblate, in the liquid state.
The deformation is stabilized by opening a gap at the Fermi level.
The closed-shell Na$_8$ remains magic also at the liquid state.

\end{abstract}


\narrowtext

Electronic shell structure of alkali metal clusters
has been established in numerous experiments (for a review
see Ref. \onlinecite{He93}). This shell structure
corresponds to single-electron levels in a spherical
potential and can be understood in terms of the simple
jellium model\cite{Br93}. It is also well-known that clusters
with an open electronic shell are deformed from the spherical
shape. In theory, the deformation has been extensively studied with
jellium-type models\cite{Cl85,La91,Ko95},
and it is also the result of ab initio
calculations\cite{Ma85,Ro91,Bo91} and simple tight-binding
models\cite{Li87,Yo94}.
These calculations seem to indicate that the geometries of smallest
clusters ($N<20$) are determined solely by the symmetries of the
single electron wavefunctions and are insensitive
to the model. Moreover, even the small
nuclei with equal number of nucleons seem to have the same geometries
as alkali metal clusters\cite{Ri68,Ko95}.

The experimental information of the deformation is obtained
from the intensity variation of the mass spectrum,
or from the photoabsorption experiments\cite{He93}
where the observed energies and the intensity ratios
of the plasmon peaks are interpreted in terms of
deformations of the valence electron distribution\cite{Se91,Bo93}.
In these experiments the clusters are known to be hot,
very likely liquidlike\cite{Ma93,Ha95}.
While the temperature can be incorporated in the electronic
degrees of freedom of the jellium model\cite{Br91},
so far
there has not been attempts
to systematically study the interplay between
the deformation of the ionic system and electronic structure
at elevated temperatures by  detailed dynamical simulations.

In this paper we  report results obtained from
ab initio  molecular
dynamics (MD) simulations of hot (500 $<$ T $<$ 1100 K )
"magic" Na$_8$ and  open-shell Na$_{14}$ clusters.
Our calculations confirm the magic nature of Na$_8$
by showing a large
 HOMO-LUMO gap  in the proposed  T=0 ground state geometry
(dodecahedron) \cite{Ro91}. This gap remains clear also at
T $\approx 550$ K, where the cluster is already liquidlike.
In the open-shell
Na$_{14}$ the deformation opens a gap,
which also remains in the liquid state.
By studying the cartesian moments of inertia, calculated from
the coordinates of ions, we show that Na$_8$
remains  the most spherical of our
clusters at high temperatures. In contrast, liquidlike
Na$_{14}$ cluster favour axially deformed (prolate and oblate)
shapes, and    the deformation of the ionic system is
driven by the opening of the HOMO-LUMO gap, which is directly
demonstrated by starting
a simulation
from a practically spherical Na$_{14}$ isomer.

Since the ab initio MD method we use is fully documented
elsewhere \cite{Ba93},  we only summarize here the key
ingredients of our calculations.
We perform the electronic-structure
calculations in the framework of the density-functional theory,
with the local density approximation for the exchange-correlation
energy. The valence electron - ion interaction is described by
 norm-conserving pseudopotentials \cite{Tr91} with a plane-wave basis.
For a given ionic configuration the Kohn-Sham one-electron equations
are solved using an iterative matrix diagonalization scheme. From
the converged solution, the Hellman-Feynman forces on the ions
are calculated, and these forces together with the classical Coulomb
forces between the positive ion cores determine the
ion trajectories.
  We remark that
due to the self-consistent
solution of the electronic structure for each ionic configuration,
the electrons are kept strictly on the Born-Oppenheimer
potential energy surface, consequently, the length of the
time step in the molecular dynamics simulation is limited only by
the vibrational time scale of ions and the performance of the
integration algorithm, just as in ordinary classical
MD.
In this study we use a fifth-order
predictor-corrector algorithm
with a time step of 5.2 fs, giving a good conservation of
the (generalized) constant of motion in the MD simulation
 \cite{Note1}.  We also remark that the method does not
 employ periodic boundary conditions to
 the ionic
system, i.e., no supercells are involved in the calculation.
For that reason the method is particularly suitable
for studies of finite charged or multipolar systems \cite{Ba93}.
The  calculations are performed typically on
a cubic cell with a dimension of 45 a.u. and using
a plane-wave kinetic energy cutoff $E_C=4.4$ Ry. Test calculations
using larger cutoffs up to 20.1 Ry show that
by using  $E_C=4.4$ Ry the
 ionization potential of a sodium atom deviates 1 \%  from
the converged value of 5.21 eV (experimental value
5.139 eV \cite{ionpot}), and the dissociation energy
and bond length of sodium dimer are within
2 \% and 5 \%,
respectively, from the converged values 0.82 eV
 and 2.98 \AA (experimental values 0.80 eV and
3.08 \AA, respectively \cite{ionpot}).

The initial configurations for the MD simulation
are prepared by setting the ions to the
desired geometry         with a reasonable nearest-neighbour distance
($\approx$ 3.5 \AA) and optimizing the structure
to a local minimum on the potential energy surface.
We wish to emphasize here that for our purposes it is
not necessary to find an absolute global ground state,
rather we wish to start the simulations from
cluster isomers with representative  electronic
properties and (in the case of Na$_{14}$)
desired
deformations in the shape of the ionic configuration.
For Na$_8$, we consider only the dodecahedron.
It has been  found to be the ground state in previous
pseudopotential-LDA calculations, where the ionic and
electronic structure are optimized according to the
Car-Parrinello scheme \cite{Ro91}.
Our optimized dodecahedron has electronic properties very similar to those
found in Ref. \onlinecite{Ro91},
 showing a large
gap of 1.1 eV    between the highest occupied and the lowest unoccupied
one-electron level.
The angular momentum
character of the four occupied levels
 is   easily identifiable showing the lowest level to be s-type
and the next three levels p-type, of which two higher
ones are degenerate. The lower p-level is split from them by
0.4 eV and the gap between the s-level and the lowest p-level is 1.2 eV.
Although a different structure (stellated tetrahedron) was
found in all-electron configuration-interaction calculations \cite{Bo91},
it has been shown \cite{Ro91} that all the low-lying isomers
of Na$_8$ have similar electronic properties, and
the cluster geometry we have selected
serves well as an example of  a  magic cluster
having a closed electronic shell.
By looking at the principal moments of inertia, we find that
two of them are degenerate and have a larger value than
the third one, thus the cluster can be characterized as
prolate,
the ratio of the two values $I_z/I_r$ ($I_r$ degenerate,
$I_z$ non-degenerate moments)
being 0.76.

We consider three isomers for Na$_{14}$, the shape of which
can be characterized as prolate, oblate and "spherical".
The prolate isomer has the lowest
energy. Its geometry was chosen
according to  a recent study
using H\"uckel method \cite{Yo94}. The cluster consists of
three stacked squares, the middle one rotated $45^\circ$ with respect
to the others. The squares are capped by an atom at both ends
of the long axis. The ratio $I_z/I_r$ is 0.39.
Of the seven occupied one-electron states,
 two lowest (identified
as s and p) and two highest (d-type, degenerate)
states   show
a single angular momentum component,
the three middle ones having some degree
of p-d mixing.
The HOMO-LUMO gap is clear, 0.4 eV.
The gap is greatly reduced for
the two other isomers
which            results in a fractional
occupancy around the Fermi level (see Ref. \onlinecite{Note1}).
The geometric structure of the oblate isomer
is obtained by relaxing two
stacked hexagonal layers of seven atoms (one atom in the middle),
rotated $30^\circ$ with respect to each other.
It is interesting to note that
the energy of this rather artificial structure is only 0.3 eV
(or 0.02 eV/atom) higher than that of the
prolate isomer. It is probable that we could find oblate
isomers which are energetically even closer
to the prolate isomer, paralleling the results of Lauritsch
et al. \cite{La91}, who find oblate and prolate shapes
(of the electron density distribution) to be
separated only by 0.1 eV in their jellium calculations
for Na$_{14}$.
Finally, the "spherical" isomer, having degenerate moments,
 is obtained by
relaxing  a structure  where an atom is placed on one (100) face
of the 13-atom cuboctahedron. The energy  of the relaxed cluster
is 0.9 eV higher       than that of  the prolate isomer.
Looking at the electronic structure we find that
the Fermi level lies in the middle of split 1d-levels, mixed with the
2s (according to the nomenclature in the spherical potential).

Next we follow the evolution of the geometric and electronic
structure of each cluster in molecular dynamics runs,
where
the cluster is
given enough kinetic energy to raise the
(vibrational)  temperature \cite{Note2} to 500-600 K, after which
the system is allowed to evolve at constant energy.
The rapid heating results in a complete disorder (melting)
during the first 1-2 ps, which is easily verified by
monitoring the mean-squared displacements (msd)  of atoms as
a function of time.
After the melting, a typical linear rise in msd versus time
is observed, and the diffusion constant estimated from the
slope of the msd curve falls in the range
$(2-9)\times 10^{-5}$ cm$^2$/s, not far from the values
$(9-13)\times 10^{-5}$ cm$^2$/s       expected for
bulk liquid sodium at these temperatures \cite{La72}.

The fluctuations of the geometric shape of the clusters
are shown in Fig. 1, which shows the evolution of the
three cartesian moments of inertia as a function of time,
and in a two-dimensional "shape space", where  a
particular time step $t_k$ is represented
by a point $(I_1(t_k)/I_3(t_k),I_2(t_k)/I_3(t_k))$,
where  $I_1<I_2<I_3$. Trajectories in this space are
confined within an upper-left triangle of a square, with
corners of (0,0), (0,1), and (1,1).
All the ideal prolate shapes $(I_1<I_2=I_3)$ fall on the line
(0,1)$\rightarrow$(1,1) while the ideal oblate
shapes $(I_1=I_2<I_3)$ form the  diagonal (0,0)$\rightarrow$(1,1)
 of the square.
    The corner (1,1) obviously
has all the shapes with
degenerate moments, i.e. shapes which have at least the
cubic symmetry.
As can be seen from Fig. 1a, the trajectory for Na$_8$ probes
relatively uniformly the region around a triaxial
average shape, characterized by coordinate (0.72,0.89).
The evolution of the
electronic structure is shown in Fig. 2a,
where we plot the Kohn-Sham eigenenergies as a
function of time.
We see only minor changes compared to zero-temperature
electronic structure: the HOMO-LUMO gap averages to
0.9 eV, and the splitting of the p-levels increases to
0.5 eV. It is interesting to note the survival of the
zero-temperature gap also between the (unoccupied)
1d-2s manifold and the 1f levels (only two of them are plotted).
This indicates that the average triaxial shape
of the ionic configuration does not destroy
too much the (approximately) spherical background potential
felt by the electrons.
The overall behaviour of the electronic shell structure
confirms the magic nature of Na$_8$ even at T=550 K,
in accordance with previous Car-Parrinello calculations \cite{Ro91}.

We turn now to the more interesting case of the open-shell
Na$_{14}$ cluster (see figs. 1b-d and 2b-d).
The results shown for the principal moments of
inertia (Fig. 1b,c) for the runs starting from the
prolate and oblate shapes show that both of these
shapes are indeed stable (meaning that there is a marked energy barrier
between them).
Furthermore, transitions between isomers
appear to be possible  in the time scale of our simulations,
$t\approx 10$ ps.
(Fig. 1b).
The gap visible at T=0 for the prolate isomer persists
also in the finite-temperature simulation (Fig. 2b),
and there is a gap opening for the oblate isomer
as soon as the dynamics is turned on (Fig. 2c).
It is instructive to note that    even the Kohn-Sham eigenvalues
reflect the shape of the cluster: for prolate shapes,
the lowest p-type orbital (corresponding
to the state along the longer axis)
is separated from the higher
p-orbitals whence for oblate shapes there is a gap between
the two lowest  (states along the two long axes)
and the higher p-orbital.  In both cases we
observe a considerable mixing between the higher p-orbital(s)
and the d-orbitals.
Finally, Fig. 2d shows a "dynamical Jahn-Teller effect" taking
place in the  "spherical" isomer of Na$_{14}$. Within 0.5 ps
after the dynamics is turned on, we see the splitting of the 1d-2s
manifold and the gap opening at $\epsilon_F$. The shape
of the cluster deforms to the prolate side (Fig. 1d),
hence also the lowest p-state splits off from the two higher p-states.

By passing we want to mention that
when the Na$_{14}$ cluster is heated to 1100 K the
time-averaged level density does not any more show a pronounced
gap at the Fermi level. At this high temperature cluster shape
changes continuously between all shapes and the
trajectory in the "shape space" covers  uniformly
a large area. It should be noted that a free cluster,
initially at 1100 K, would very fast cool by evaporation,
though evaporation is not observed in our simulations
due to the short time scale.
Thus it is likely that at experimental temperatures
the Na$_{14}$ cluster is strongly deformed also in the liquid state.

In conclusion,
we have compared the behaviour of hot "magic" Na$_8$ and open-shell
Na$_{14}$ clusters in ab initio molecular dynamics simulations
at elevated vibrational temperatures from 500 to 1100 K,
where the ionic structure
of all the clusters is completely disordered and exhibits
liquidlike diffusion.
Na$_8$ remains magic around 500 K  showing a clear electronic
shell structure. We find two stable shape isomers (prolate and oblate)
for Na$_{14}$ and observe inter-isomeric transitions
in our 10 ps time scale for clusters around
600 K.
We demonstrate a "dynamical Jahn-Teller effect" in a spherical
isomer of Na$_{14}$ where the Fermi level initially lies in the
middle of the 1d-2s manifold, but once the dynamics is turned on,
the cluster  opens a gap
at $\epsilon_F$, which stabilizes the deformation
to the prolate shape.

\begin{acknowledgments}

We wish to thank Robert N. Barnett and Uzi Landman at Georgia
Institute of Technology, Atlanta, for illuminating discussions
on the ab initio MD method and for sharing the computer codes.
This study is supported by the Academy of Finland.
Calculations have been performed at the Center for Scientific Computing,
Espoo, Finland, and at the Florida State University
Computing Center.
\end{acknowledgments}

\begin{figure}
\caption{Time evolution of the cartesian
 moments of inertia of liquid sodium clusters.
(a) Na$_8$ at 550 K, (b) prolate Na$_{14}$ at 650 K, (c) oblate
Na$_{14}$ at 610 K, (d) Na$_{14}$ at 480 K started from a spherical shape.
For each cluster we show the trajectory in the phase space and the three
moments of inertia  (in units of 10$^3$ (amu)a$_0^2$)
as a function of the time in ps (inset). Note the transition from the
prolate shape to oblate shape (and back) in (b), and the
rapid strong deformation in (d).}
\end{figure}

\begin{figure}
\caption{Evolution of single-electron eigenvalues (in eV)
as a function of time  in ps (left)
and the time-averaged density of states  (in arbitrary
units, right) in
(a) Na$_8$ at 550 K, (b) prolate Na$_{14}$ at 650 K, (c) oblate
Na$_{14}$ at 610 K, (d) Na$_{14}$ at 480 K started from a spherical shape.
Note the gap opening at $\epsilon_F$ in (d).}
\end{figure}


\begin{references}

\bibitem{He93} W. A. de Heer, Rev. Mod. Phys. {\bf 65}, 611 (1993).

\bibitem{Br93} M. Brack, Rev. Mod. Phys. {\bf 65}, 677 (1993).

\bibitem{Cl85} K. Clemenger, Phys. Rev. B {\bf 32}, 1359 (1985);
 W. Ekardt and Z. Penzar, Phys. Rev. B {\bf 43}, 1322 (1991);
 S. Frauendorf and V. V. Pashkevich, Z. Phys. D {\bf 26}, S98 (1993);
 A. Bulgac and C. Lewenkopf, Phys. Rev. Lett. {\bf 71}, 4130 (1993);
 C. Yannouleas and U. Landman, Phys. Rev. B {\bf 51}, 1902 (1995);
 B. Montag, Th. Hirschmann, J. Meyer, P.-G. Reinhard, and M. Brack,
 to be published.

\bibitem{La91} G. Lauritsch, P.-G. Reinhard, J. Meyer, and M. Brack,
Phys. Lett. A {\bf 160}, 179 (1991).

\bibitem{Ko95} M. Koskinen, P.O. Lipas, and M. Manninen, to be published.

\bibitem{Ma85} J. Martins, J. Buttet, and R. Car, Phys. Rev. B {\bf 31},
1804 (1985).

\bibitem{Ro91} U. R\"othlisberger and W. Andreoni,
J. Chem. Phys. {\bf 94}, 8129   (1991).


\bibitem{Bo91} V. Bona\v ci\v c-Kouteck\'y, P. Fantucci, and J. Kouteck\'y,
Chem. Rev. {\bf 91}, 1035 (1991).

\bibitem{Li87} D. M. Lindsay, Y. Wang, and T. F. George, J. Chem. Phys.
{\bf 86}, 3500 (1987).

\bibitem{Yo94} A. Yoshida, T. D\o ssing and M. Manninen, J. Chem. Phys.
{\bf 101},  3041 (1994).

\bibitem{Ri68} G. Ripka, Adv. Nucl. Phys. {\bf 1}, 183 (1968).

\bibitem{Se91} K. Selby, V. Kresin, J. Masui, M. Vollmer, W. A. de Heer,
A. Scheidemann, and W. Knight, Phys. Rev. B {\bf 43}, 4565 (1991).

\bibitem{Bo93} J. Borggreen, P. Chowdhury, N. Kebaili, L. Lundsberg-Nielsen,
K. L\"utzenkirchen, M. B. Nielsen, J. Pedersen and H. D. Rasmussen,
Phys. Rev. B {\bf 48}, 17507 (1993).

\bibitem{Ma93} T.P. Martin, U. N\"aher, H. Schaber, and U. Zimmermann,
J. Chem. Phys. {\bf 100}, 2322 (1993).

\bibitem{Ha95} K. Hansen and J. Falk, to be published.

\bibitem{Br91} M. Brack, O. Genzken, and K. Hansen, Z. Phys. D {\bf 21},
65 (1991).

\bibitem{Ba93} For the method, see
R. N. Barnett and U. Landman, Phys. Rev. B {\bf 48}, 2081 (1993),
and for recent applications, see Refs. 28-31 therein and Ref. \onlinecite{Ha94}
of this paper.

\bibitem{Note1} Since we allow fractional occupation numbers
calculated from a Fermi distribution (with a preassigned
electronic temperature  $T_{el}\approx 300$ K)
in situations where the Kohn-Sham eigenstates are (nearly)
degenerate at the Fermi surface (Ref. \onlinecite{Ba93}),
the conserved quantity in the MD simulation is the
free energy
$\Omega=U-T_{el}S$, where $U$ is the total internal
energy of the electron-gas system, and
$S$ is the entropy of the non-interacting electron gas.
$\lbrack$See e.g. R. M. Wentzcovitch, et. al., Phys. Rev. B
{\bf 45}, 11372 (1992).$\rbrack$
The relative fluctuations of
$\Omega$ are smaller than 0.003\%
 (at $T\approx 600$ K) throughout
our simulations, when
a time step of 5.2 fs is used.

\bibitem{Tr91} N. Troullier and J. L. Martins, Phys. Rev. B {\bf 43},
1993 (1991). The pseudopotential for sodium has been generated
and tested by us previously   in simulations of sodium chloride
clusters, see Ref. \onlinecite{Ha94}.

\bibitem{Ha94} H. H\"akkinen, R. N. Barnett and U. Landman,
Europhys. Lett. {\bf 28}, 263 (1994); Chem. Phys. Lett. {\bf 232},
79 (1995).

\bibitem{ionpot} {\it CRC Handbook of Chemistry and Physics},
ed. R.C. Weast (CRC Press, Boca Raton 1986);
K. P. Huber and G. Herzberg, {\it Constants of
Diatomic Molecules} (Van Nostrand Reinhold, New York 1979).


\bibitem{Note2} The initial velocities to ions are given such that
the components of linear and angular momenta are zeroed. Hence,
we calculate the temperature of the cluster as
$T=2E_k/(3N-6)k_B$, where $E_k$ is the total kinetic energy of ions,
$N$ number of ions, and $k_B$ Boltzman constant.

\bibitem{La72} S. J. Larsson, C. Roxbergh, and A. Lodding, Phys. Chem.
Liq. {\bf 3}, 137 (1972); G.-X. Qian, M. Weinert, G. W. Fernando,
and J. W. Davenport, Phys. Rev. Lett. {\bf 64}, 1146 (1990).

\end{references}
\end{document}